\documentclass[amstex,times,11pt,twoside]{acta}
\usepackage{amssymb}
\usepackage{amsmath}
\SetPages{000}{000}
\SetVol{50}{2000}
\usepackage{graphics}

\begin{document}


\begin{Titlepage}
  \Title{Radiation Spectra of Advection Dominated Accretion Flows around
         Kerr Black Holes}

  \Author{ A. Kurpiewski 
           \and M. Jaroszy\'nski }
         {Warsaw University Observatory, Al.~Ujazdowskie~4,~00-478~Warszawa,
e-mail:kurpiew@astrouw.edu.pl,mj@astrouw.edu.pl}
\vspace*{9pt}

\Received{2000}

\end{Titlepage}

\Abstract{
We study the formation of the spectra in advection dominated accretion
flows (ADAFs) around Kerr black holes. We use a Monte Carlo approach and
fully general relativistic treatment to follow the paths of individual
photons and model their scattering with mildly relativistic, thermal
electrons of the two temperature plasma present in the flow. We 
study the influence of the accretion rate, black hole mass, black hole 
angular momentum, and the strength of the small scale magnetic field
present in the plasma on the resulting spectra. The impact of the black
hole angular momentum on the spectra is present and can be distinguished
from the influence of changes in other parameters.
This property of the models should be taken
into account when modeling the individual sources and the population of
inefficiently accreting black holes in the Universe.
}
{black holes - accretion - radiation processes}

\section{Introduction}

The Advection Dominated Accretion Flows (ADAFs) have been reviewed
by a number of authors, most recently by 
Yi (1999), Lasota (1999),
Narayan, Mahadevan and Quataert (1998),
Svensson (1998),
and Kato, Fukue and Mineshige (1998).
The ADAFs represent a class of optically thin solutions of
the accretion flow, which radiate so inefficiently, that almost all the
heat dissipated inside the fluid is subsequently advected toward the black
hole horizon. Since the cooling of matter is negligible, the equations of
fluid dynamics are independent of the equations describing the emission,
absorption and scattering of radiation. This allows one to address these two
topics separately. In this paper we are concerned mostly with the radiation
processes inside the flow.

The fully relativistic dynamics of ADAFs has been described and some
solutions have been obtained in Lasota (1994),
Abramowicz et al. (1996),
Abra\-mo\-wicz, Lanza and Percival (1997),
Peitz and Appl (1997),
Jaroszy\'{n}ski and Kurpiewski (1997, hereafter Paper I), 
Gammie and Popham (1998),
and Popham and Gammie (1998).
The most extensive survey of the parameter
space is probably presented by Popham and Gammie (1998),
where the dependence of the flow on the black hole spin defined by the
dimensionless Kerr parameter $a$, 
the viscosity
parameter $\alpha$, the gas adiabatic index, and on the advection parameter
$f$ (where $f=1$ means fully advective flow without any cooling) is
investigated. This work shows rather strong dependence of the flow
characteristics on parameters mentioned, especially on the black hole spin.
The dependence on advection parameter is also substantial, but for a narrow
range of $f$, which can represent flows with negligible cooling 
($0.9 \le f \le 1$) one can use the $f=1$ solutions. 

In our calculations we use the solutions of the equations describing the
flow dynamics presented in Paper I. 
All our models use $f=1$, which is representative of the physically 
relevant ADAFs. The black hole spin, we
consider, is limited to the three values ($a=0$, $0.5$, and $0.9$). 
We also use only one value of the viscosity parameter $\alpha=0.1$. 

The main purpose of this paper is a self-consistent
treatment of photon Comp\-to\-ni\-za\-tion in a two temperature plasma of
an ADAF solution and a limited survey of the parameter space of the
models. We check the dependence of the resulting spectra on the black
hole mass and spin, accretion rate, and the
parameter describing the importance of the small scale magnetic fields
$\beta$. 

In our previous paper on the subject (Kurpiewski and Jaroszyñski 1999,
hereafter Paper II) we have described the methods of calculating the
spectra of ADAFs using Monte Carlo approach to follow the individual
photons and taking into account all the relativistic effects in photon
propagation. We use this approach here with two changes: following
Nakamura et al. (1997) and Narayan et al. (1998, hereafter NMGPG) we
modify the thermal balance equation adding terms describing viscous
heating of the electrons and advection of heat by them. 

In the next Section we present the modifications of the methods employed
in Papers I and II.  In Sec.~3 we present the results of calculations, 
showing some details of the structure of the various ADAF configurations
and the spectra produced by them.
The discussion and conclusions follow in the last Section.

\section{The model}

The dynamics of the flow, the methods of calculating the cooling rate at
a chosen location in the configuration when the density and electron
temperature  are given, as well as the Monte Carlo approach to
Comptonization are fully described in Papers I and II. We are not going
to repeat it here. The only substantial change in our methods compared
to the previous papers is the new treatment of the thermal equilibrium
in the two temperature plasma. Following Nakamura et al. (1997) and
NMGPG we calculate the amount of heat advected by the
electron gas. This factor is unimportant in the energy budget of the
whole configuration but may be important in finding the temperature of
the electrons. We also include the possibility that the electrons gain 
some thermal energy directly from the dissipative processes.
Both processes influence the electron thermal energy balance.
Such terms have been usually neglected under the assumption that the
equilibrium between Coulomb heating and various cooling processes for
electrons is quickly established.

Our models are obtained under many simplifying assumptions. In
particular we solve the equations for the ``vertically averaged, stationary
configuration" (Abramowicz et al. 1996; Narayan and Yi 1994) which is a
standard procedure in ADAFs. We make also some more detailed assumptions
postulating that in the spherical coordinate system (Boyer Lindquist
coordinates, see Bardeen 1973) 
the fluid has no poloidal velocity component. We also
assume that the speed of sound, the radial velocity component, and the
specific angular momentum of the fluid depend only on the radial
coordinate. These assumptions are sufficient to fully describe the
pressure and density distribution in three dimensional configuration
(see Paper II). The distribution of electron temperature needs further
treatment which we describe with some details below.

The thermodynamics of the gas and magnetic field is described by the set
of equations given by NMGPG. We only introduce some
factors which one may treat as relativistic corrections. 

In ADAF models which use synchrotron radiation as a source of soft photons
it is usually assumed that the ratio of the magnetic pressure to the gas
pressure is constant ($P_{\mathrm m}/P_{\mathrm g}=(1-\beta)/\beta$ where 
$P_{\mathrm g}=\beta P_{{\mathrm{tot}}}$, 
$P_{\mathrm m}=(1-\beta)P_{\mathrm{tot}}$).
NMGPG go further and assume that the magnetic field is separately
"assigned" to electrons and ions, and the the ratio of gas to magnetic
pressure in each component is the same. That implies that also some part
of magnetic field energy is associated with the electron gas:
  \begin{equation}
    P_{\mathrm e}={1 \over \beta}n_{\mathrm e}kT_{\mathrm e}~~~~
    U_{\mathrm e}=\left[{3(1-\beta) \over \beta} + a(T_{\mathrm e})\right]
    {n_{\mathrm e}kT_{\mathrm e} \over \rho}
  \end{equation}
where $n_{\mathrm e}$ is the electron concentration, $T_{\mathrm e}$ is
the temperature of the electron gas, $k$ is the Boltzmann constant, and 
$U_{\mathrm e}$ is the energy density associated with electrons per unit
rest mass of the fluid with rest mass density $\rho$.
The ratio of the energy density to the pressure for the electron
component of the fluid $a(T_{\mathrm e})$ is given as:
  \begin{equation}
    a(T_{\mathrm e})={1 \over \Theta_{\mathrm e}}
    \left[{3{\mathrm K}_3(1/\Theta_{\mathrm e}) 
    +{\mathrm K}_1(1/\Theta_{\mathrm e}) \over
    4{\mathrm K}_3(1/\Theta_{\mathrm e}) }   -1\right]
  \end{equation}
In the above formula we use the dimensionless electron temperature 
$\Theta_{\mathrm e}=kT_{\mathrm e}/m_{\mathrm e}c^2$, and ${\mathrm K}_i$ are
the modified Bessel functions. 

The term describing the advection of heat by electrons 
$q_\mathrm{adv,e}$  is proportional to
the entropy gradient along the streamlines, and can be expressed using
the temperature and density gradients (NMGPG):
  \begin{equation}
    q_\mathrm{adv,e} =
    n_{\mathrm e}kT_{\mathrm e}u^r\left[
    \left({3-3\beta \over \beta}+a(T_{\mathrm e})
    +T_{\mathrm e}{{\mathrm d}a \over {\mathrm d}T_{\mathrm e}}\right)
    {{\mathrm d}T_{\mathrm e} \over T_{\mathrm e}{\mathrm d}r}
    -{1 \over \beta \rho}{{\mathrm d}\rho \over {\mathrm d}r}\right]
  \end{equation}
where $u^r$ is the radial component of the four velocity in the Boyer
Lindquist coordinates.
The equation of the thermal equilibrium for electrons has the form
  \begin{equation}
    q_{\mathrm{ie}}^{+}+q_{\mathrm{vis,e}}^{+} = q_\mathrm{adv,e}+
    q_{\mathrm{s,C}}^{-}+q_{\mathrm{br,C}}^{-}
  \end{equation}
where $q_{\mathrm{ie}}^+$ is the rate of heating of electrons by Coulomb
interactions with ions, 
$q_{\mathrm{vis,e}}^+$ is the rate of the direct heating of electrons by
dissipative processes,
$q_{\mathrm{s,C}}^-$ is the synchrotron,
and $q_{\mathrm{br,C}}^{-}$ the bremsstrahlung cooling rate, both including
Comptonization. 
We assume that the direct energy dissipation to the electron gas is a
small fraction $\delta$ of the total dissipation rate 
$q_{\mathrm{vis,e}}^{+}=\delta q_{\mathrm{vis}}^{+}$. Methods of calculating
terms $q_{\mathrm{ie}}^{+}$
$q_{\mathrm{vis}}^{+}$,
$q_{\mathrm{s,C}}^{-}$,
and $q_{\mathrm{br,C}}^{-}$ are given in Papers I and II.

We find the distribution of electron temperature in the configuration
iteratively solving the equation for thermal balance. Because the advection
of heat by electron gas is now included in the equations, the conditions in
any place in the disk depend on the conditions upstream. In our
iterations we start from the outer layers of the configuration and
follow the fluid inward. The temperature of
the electrons in the whole configuration has an impact on the average
optical depth, especially for soft photons.
This influences the characteristic frequency of synchrotron
selfabsorption, and hence the rate of cooling. 
The approach to this problem of non-locality has been described in
Paper II. As our previous calculations show, the electron tempe\-rature in a
given location does not depend strongly on the conditions around. Thus it
is possible to obtain a selfconsistent electron temperature distribution
after a few iterations, even if one starts from a primitive first
approximation $T_{\mathrm e}(r,\theta)={\mathrm{const}}$.


\section{Results}

We have calculated several ADAF models with parameters listed in Table 1. 
These parameters are: the
mass of the black hole expressed in solar masses $m=M_\mathrm{BH}/M_\odot$,
the accretion rate in units of the critical accretion rate $\dot{m} =
\dot{M}/\dot{M}_\mathrm{crit}$ (where $\dot{M}_\mathrm{crit} =
L_\mathrm{Edd}/c^2 = 1.4 \times 10^{17} m~~\mathrm{g}~\mathrm{s}^{-1}$) 
and parameters $\beta$ and $\delta$. The black hole spin $a$ takes three
values: 0, 0.5 and 0.9 in each model. We have chosen only one value of 
the viscosity parameter, $\alpha = 0.1$, common for all models. 
We use gravitational radius $r_\mathrm{g}=GM/c^2$, where $G$ is the
gravity constant and $c$ is the speed of light.
We have 
assumed that at the outer boundary of the flow $r_\mathrm{out} = 10^3 
r_\mathrm{g}$ the specific angular momentum is close to 95\% of
the Keplerian specific angular momentum ($\ell \approx
0.95~\ell_\mathrm{K}$). The radial component of the velocity is much
smaller than the speed of sound at the boundary, so the flow is subsonic
at the outer boundary.
In Models 9 and 10 we have chosen the same values of $m$, $\dot{m}$ and 
$\beta$ as in Paper II to examine the influence of the new heating and 
cooling rates in the thermal equilibrium equation (4) on the radiation 
spectra of the flow. 

  \begin{table}[h]
  \begin{center}
  \caption{\small Parameters of models.}
  \begin{tabular}{ccccc}
    \hline   
     Model & m & $\dot{m}$ & $\beta$ & $\delta$ \\
    \hline
     1  & $10^8$ & $10^{-2}$ & 0.95 & $10^{-3}$ \\
     2  & $10^8$ & $10^{-2}$ & 0.5  & $10^{-3}$ \\
     3  & $10^8$ & $10^{-3}$ & 0.95 & $10^{-3}$ \\
     4  & $10^8$ & $10^{-3}$ & 0.5  & $10^{-3}$ \\
     5  & $10^7$ & $10^{-2}$ & 0.95 & $10^{-3}$ \\
     6  & $10^7$ & $10^{-2}$ & 0.5  & $10^{-3}$ \\
     7  & $10^7$ & $10^{-3}$ & 0.95 & $10^{-3}$ \\
     8  & $10^7$ & $10^{-3}$ & 0.5  & $10^{-3}$ \\
    \hline
     9  & $3.6\times10^7$ & 0.016 & 0.95 & $10^{-3}$ \\
     10 & $3.6\times10^7$ & 0.016 & 0.95 & 0         \\
    \hline
  \end{tabular}
  \end{center}
  \end{table}

As in Paper II we solve the equations for thermal equilibrium 
(4) and the equation of state on the 2-dimensional grid  in the 
($r,\theta$) plane. 
The grid cells are spaced logarithmically in radius and linearly in 
polar angle. We use 25 values of radius 
$\lg r \in [\lg r_\mathrm{h}+0.05;3]$, where $r_\mathrm{h} =
(1+\sqrt{1-a^2})r_\mathrm{g}$ is the radius of the black hole event
horizon. The polar angle $\theta$ (measured from the rotation axis) 
takes 19 values from $0^{\circ}$ to $90^{\circ}$ with $\Delta \theta = 
5^{\circ}$. The radiation spectra of the flow are calculated in the same
manner as in Paper II. Using Monte Carlo methods we have obtained
Comptonized spectra of input synchrotron and bremsstrahlung photons in
function of their frequency at infinity. We have also investigated the
dependence of the observed flux of radiation on the inclination angle 
of the observer relative to the axis of rotation. The
resulting spectra and bolometric fluxes presented in this paper are
based on calculations including $3\times 10^5$ synchrotron and $3\times
10^5$ bremsstrahlung input photons and following about 3 - 5 times as many
branches of photon trajectories in each case. 

\begin{figure}[h]
\vspace{10.5cm}
\includegraphics{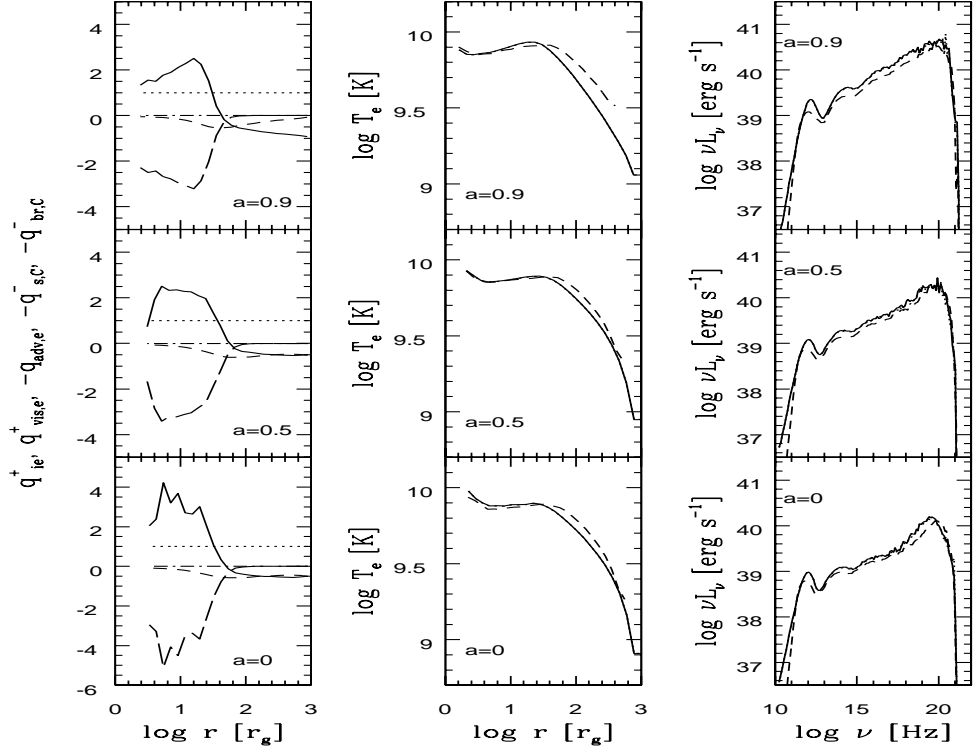}
\caption{\small The influence of the direct electron heating and
advection of heat by electrons on their temperature distribution
and on the resulting spectra. For comparison we show the results for the
models with parameters  
($m$,$\dot{m}$,$\beta$) = ($3.6\times10^7$,$0.016$,$0.95$) of Paper II.
{\bf Left panel:}
The relative importance of various heating (positive values) 
and cooling (negative values) processes at the equator for Model 9. 
The rate of electron heating by Coulomb interactions with ions 
$q_{\mathrm{ie}}^{+}$ is used for comparison and shown as dotted
straight line at value 1. 
The rate of the direct heating by dissipative processes
$q_{\mathrm{vis,e}}^+$ (dot-dashed),
the term describing the advection of heat by 
electrons $q_{\mathrm{adv,e}}$ (solid),
the synchrotron cooling rate including Comptonization $q_{\mathrm{s,C}}^-$
(long-dashed),
and the bremsstrahlung cooling rate including Comptonization 
$q_{\mathrm{br,C}}^{-}$ (short-dashed) are shown as functions of radius
in units of $q_\mathrm{ie}^+(r)$.
{\bf Middle panel:} The radial dependence of
the electron temperature at the equator for Model 9 (solid lines), Model
10 (dotted lines, covered with solid lines) and model described in 
Paper II (dashed lines). {\bf Right panel:} The radiation spectra of ADAF
for these cases. The three diagrams in each panel correspond to different
values of the black hole spin, as indicated.}
\end{figure}

\begin{figure}[h]
\vspace{16cm}
\includegraphics{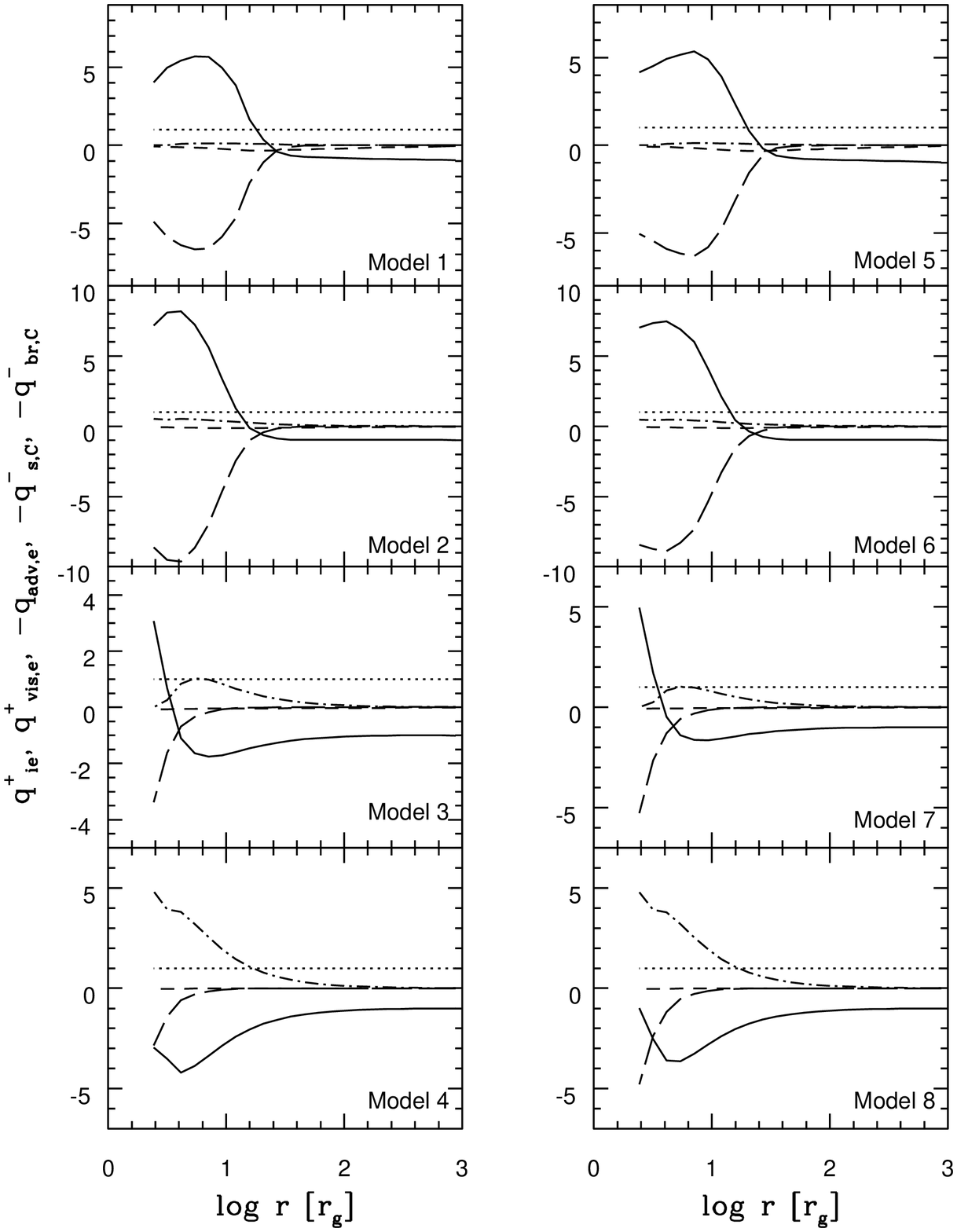}
\caption{\small The same as in left panel of Fig. 1, but for Models 1-8.}
\end{figure}

First, we examine the impact of the new terms in the thermal
balance equation  (the direct heating of electrons by dissipative 
processes $q_\mathrm{vis,e}^{+}$ and the advection of
heat energy by electrons $q_\mathrm{adv,e}$) on the radial 
distribution of the electron tempe\-rature and radiation spectra of the 
flow. In the middle and right panels of Fig. 1 we compare the results
obtained in Model 9 (electron advection and dissipation included), Model
10 (dissipation neglected) and the model from Paper II (both neglected).
It is clearly seen that the direct energy dissipation to the electrons 
does not affect the energetic balance of the electron gas,
so it can not influence the radiation spectrum. The spectra with
advection by electrons and without it differ noticeably but not
substantially. 

In the left panel of Fig. 1 we show the radial dependence of the cooling
and heating rates of electrons in Model 9. 
The advection of heat by electrons is 
the efficient cooling process in the outer part of the flow
($\lg r \gtrsim 1.5$) and the efficient heating process in the
inner part of the flow. It causes changes in the radial distribution of
the electron temperature but not large and mostly in the outer region
($\lg r \gtrsim 1.5$). The changes in the radiation spectra
(especially in the slope of Comptonized spectra of synchrotron radiation)
are also not big.

\begin{figure}[h]
\vspace{16cm}
\includegraphics{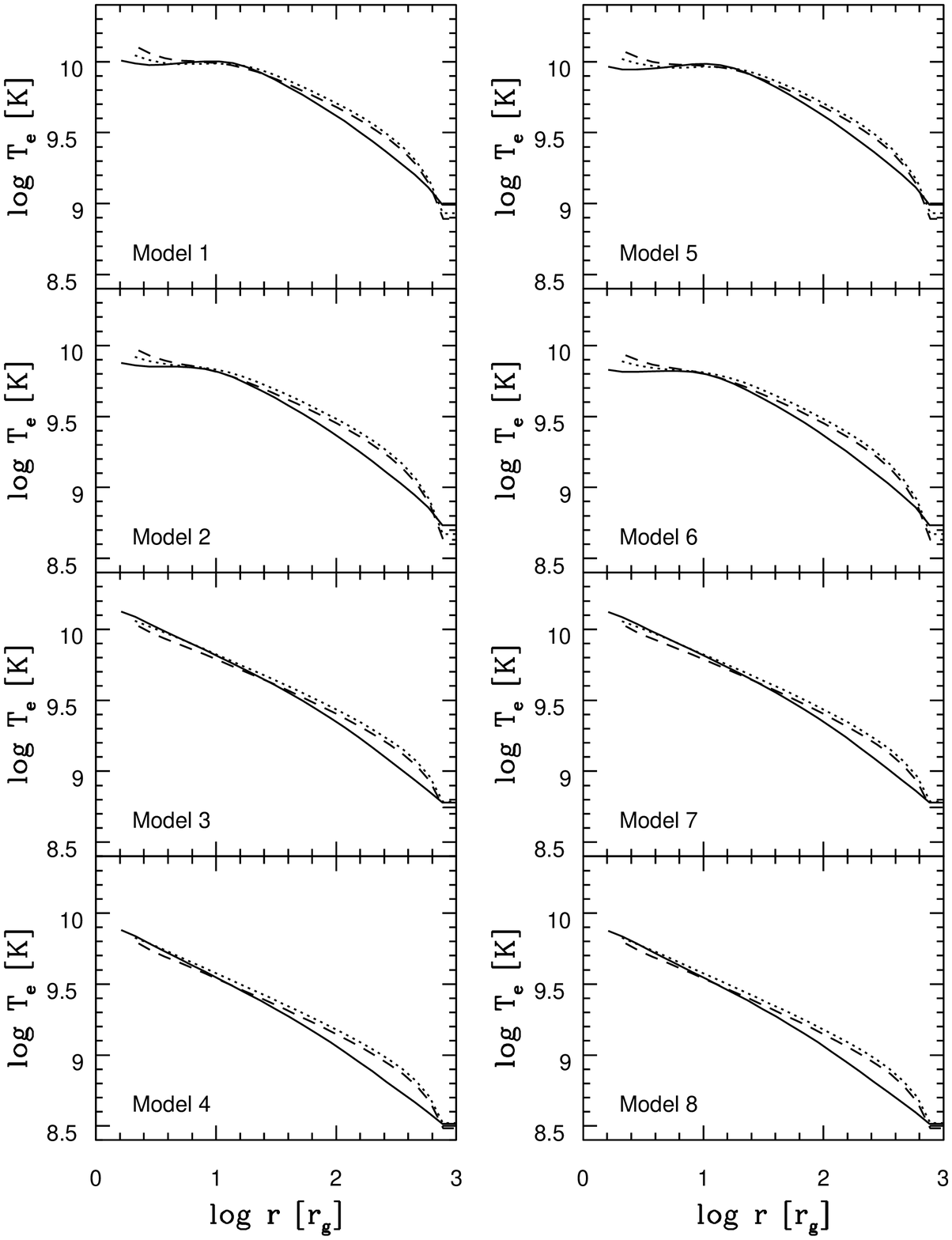}
\caption{\small The radial dependence of the electron temperature at the
equator for Models 1-8. The short-dashed, dotted and solid lines
represent
the cases with a=0, 0.5, 0.9, respectively.}
\end{figure}

\begin{figure}[h] 
\vspace{16cm}
\includegraphics{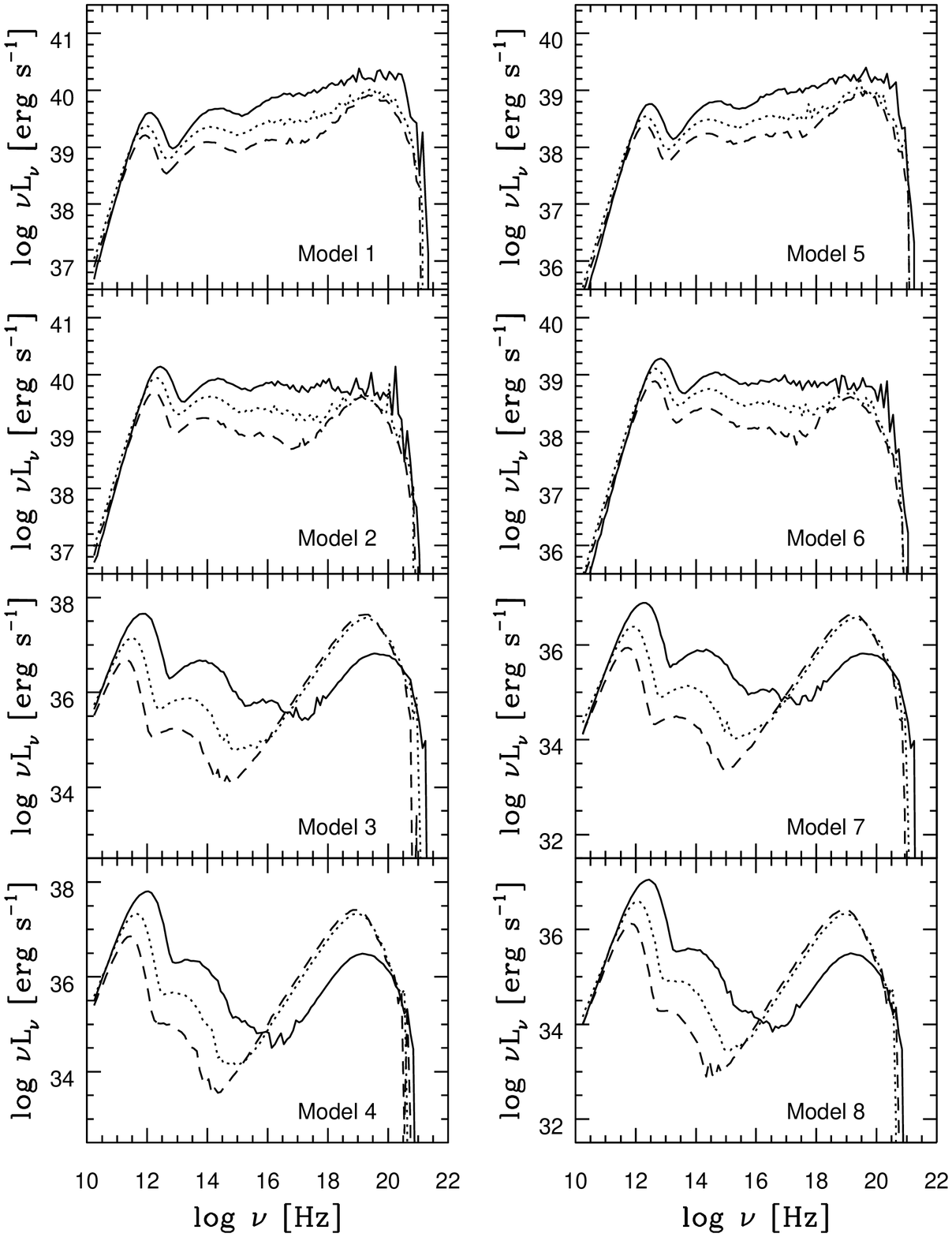}
\caption{\small The radiation spectra of ADAF for Models 1-8. The three
cases of the Kerr parameter $a$ are represented by the same types of
lines as in the Fig. 3.}
\end{figure}

Figure 2 shows the effects of model parameters $m$, $\dot{m}$ and
$\beta$ on the radial distribution of the cooling and heating rates of
electrons. We present the results for a=0.9 only, because the influence
of the black hole spin is not big. We see that also the black hole mass
does not affect strongly the energetic balance of electrons. Much more
important are the accretion rate and parameter $\beta$. The smaller are 
$\dot{m}$ and $\beta$ (it means smaller density of the flow and
higher density of the magnetic field in the plasma), the stronger is 
the influence of the direct energy dissipation to the electrons
on their energetics. Simultaneously  the radius at which advection by
electrons becomes a heating process decreases.  For $\dot{m} =
10^{-3}$ and $\beta = 0.5$ (Models 4 and 8), for example, there is no 
advective heating of the electrons -- all the advectively transported 
heat is carried in under the black hole horizon. These effects cause the
changes in radial distribution of the electron temperature in the flow,
which is shown in Fig. 3. With decreasing $\dot{m}$ and $\beta$ the
profile of the radial distribution of the electron temperature is more
and more steep losing the characteristic flat nature for small radii. 
These results are in good agreement with results of Nakamura et al.
(1997) and NMGPG.

Finally, in Fig. 4 and Fig. 5 we show the influence of the model
parameters on the radiation spectra $\nu L_\nu$ and on the observer -
inclination dependence of the bolometric flux measured at infinity, $L(i)$. 
There are three main components of the radiation
spectra of ADAFs which dominate in the $\lg(\nu)$ vs. $\lg(\nu L_\nu)$
plot: the synchrotron peak (radio energies), the
bremsstrahlung peak (X-ray and soft $\gamma$-ray energies) and the
Comptonized spectra of the synchrotron radiation between them (sometimes
this component covers the bremsstrahlung peak). The shapes and 
relative positions of the synchrotron and bremsstrahlung peaks in the plot,
as well as the slope of the Comptonized spectrum (expressed by power law index 
$\Gamma$, where $L_\nu \sim \nu^{-\Gamma}$) depend on the distribution of the
electron temperature and density in the flow, so they depend on the
model parameters. 

\begin{figure}[h]
\vspace{16cm}
\includegraphics{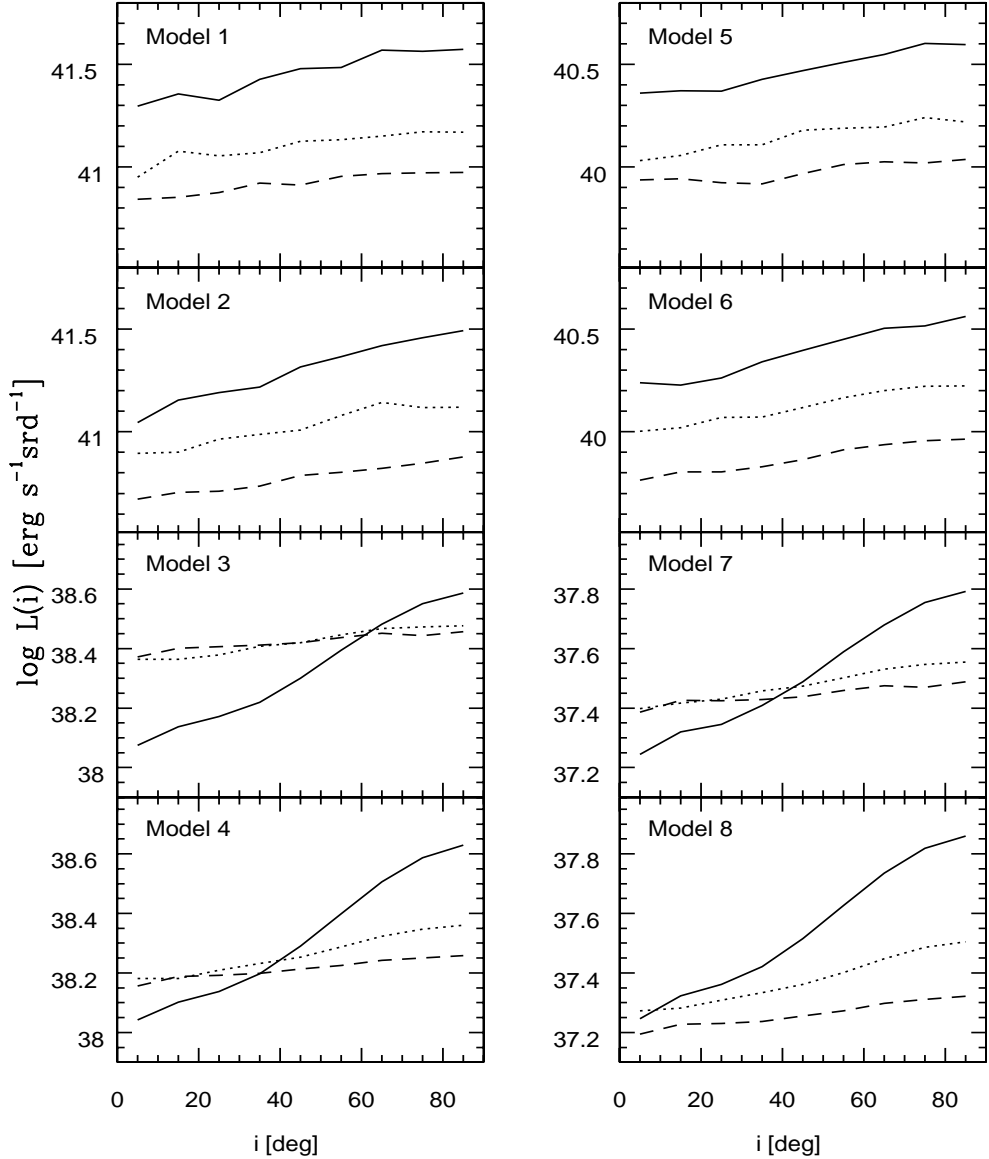} 
\caption{\small The dependence of the observed total bolometric flux
from the ADAF on the observer inclination relative to the axis of
rotation for Models 1-8. The three cases of the black hole spin $a$
are represented by the same types of lines as in the Fig. 3.}
\end{figure}

There is very weak dependence of the radiation spectra and luminosity
aniso\-tropy on the black hole mass. The total bolometric luminosity, however,
is directly proportional to $m$. With decreasing $\beta$ the synchrotron
peak becomes higher, the power law index of the Comptonized spectra
becomes smaller and $L(i)$ slightly steeper. The greatest influence on
radiation spectra and bolometric flux has the accretion rate. This
topic, however, has been examined very well in many papers devoted to 
ADAFs and the dependencies of bolometric luminosity of each component of
the ADAFs spectrum on $\dot{m}$ is rather obvious. The most interesting
result of our investigations in this paper (and Paper II) is the 
influence of the black hole angular momentum on radiation spectra.
With increasing $a$ the index $\Gamma$ becomes lower which corresponds
to flatter $L_\nu \sim \nu^{-\Gamma}$ spectrum with relatively more
high energy photons. The total luminosity becomes higher and the  anisotropy 
of the source is more  pronounced, which corresponds to the steeper 
dependence of the observed luminosity on the inclination angle,  $L(i)$.
The anisotropy is not very big. In most cases the ratio of the
luminosity observed at the equator to the luminosity observed on the
rotation axis $L(90^\circ)/L(0^\circ) \in [1.1; 2]$.
The exceptions are the models with $a=0.9$ and $\dot{m} = 10^{-3}$, where
this ratio reaches values $\sim 4$.

\section{Discussion and conclusions}

In this paper we have used the dynamical models of ADAFs obtained with
the methods of Papers I and II. We assume that the influence of the cooling
processes on the dynamics of the flow can be neglected, since the
advection of heat dominates. This assumption allows us to 
solve the problem of heat and radiation transport separately. 
We do not consider the possibility of matter outflow from the inner
parts of the disk. This scenario, called Advection Dominated Inflow -
Outflow Solutions,  has been put forward by Blandford and Begelman (1999).
Following the discussion presented by Paczyñski (1998) based on his
semi-analytic model of the flow and also the physical  and 2D numerical
analysis of the problem by  Abramowicz, Lasota and Igumenshchev (2000) we
think that models neglecting the outflows are selfconsistent. 

The outer boundary conditions in our models resemble the {\it Preheated}
ADAFs of Park and Ostriker (1999; 2000) in the sense that the thermal
energy of the fluid at the outer edge of the disk is a substantial 
part of its virial energy.  The problem of transition from Keplerian 
disk to advection dominated flow has not been solved yet so the outer 
boundary condition can not be introduced in a "natural" way. 
Other aspects of our method are different: we use solutions of height
averaged equations for the disk structure instead of self-similar solutions
of Narayan and Yi (1995) employed by Park and Ostriker (2000). To get the
two dimensional distribution of the fluid density we use the approach
of Paper II. We assume the specific angular momentum  and the radial
component of the velocity to be constant on $r=\mathrm{const}$
"spheres". This is sufficient to calculate the density distribution on
the spheres. In our approach the empty funnels near the rotation axis 
arise naturally due to the existence of centrifugal potential barrier. 
The funnels are not conical and their opening angles vary from $\sim
30^\circ$  at small radii to $\sim 5^\circ$ at larger distances from
the black hole. These results are in agreement with Park and Ostriker
(2000) calculations assuming preheating of ADAFs.

We have not explored the models for as large a range of accretion rates as
has been done by Park and Ostriker (2000), so we can not comment on the 
limits on accretion rate ${\dot m}$ for which the models remain
selfconsistent  and belong to the class of (Preheated) ADAFs. In our
calculations we have not met any problems with energy balance equation.
We think that conclusions referring to the necessity of including matter
outflows and/or preheating in modeling ADAFs must await a fully 2D
treatment of the combined dynamical and radiative transfer equations. 

In this paper we have changed our equation for the thermal energy balance
of Paper II including the 
direct heating of electrons by dissipation and taking into account the
advection of heat by electron gas (NMGPG; Nakamura et al. 1997). The
direct comparison of the models including and not including the changes
shows, that the heating of electrons by dissipation (at the rate $10^3$
times lower than for ions) has practically no influence on the electron
temperature distribution and does not change the resulting radiation
spectra of the flows. On the contrary, the advection of heat by
electrons substantially changes the thermal structure of the flow. The
advection plays the role of cooling process in the outer regions of the
disks, where most of the energy delivered to the electrons by Coulomb
interactions is stored and transported inward.  In the inner parts of
the flow some part of this energy is radiated, especially in high
accretion rate cases (${\dot m} \gtrsim 10^{-2}$).

We have calculated a larger family of models using slightly improved methods
as compared with Paper II. Our calculations include all relativistic
effects in light propagation and scattering and employ extensively Monte
Carlo approach.  Our conclusion from Paper II remains valid: for other
parameters kept constant, the spectra of more rapidly rotating black holes
are harder. The Doppler boost of photons, which happen to be scattered many
times,  is probably one of the factors, but more important factor is the 
systematic difference between density distribution in the inner parts of
the flow depending on the angular momentum of the hole. For spinning
holes the sonic radius is smaller. Inside this radius the flow
is close to the free fall, but outside it the pressure gradients
partially control and slow down the radial velocity of the flow. At
given radius the velocity of inflow toward the rapidly rotating black
hole is slower than the velocity toward a Schwarzschild black hole,
because the distance remaining to the sonic point is larger in this
case. If the accretion rates in both cases are the same, the density
must be higher in the case of rotating hole. Higher density explains
relatively higher rate of synchrotron radiation and a higher probability of
scattering. Such effects are seen in our spectra: the first synchrotron
peaks have heights monotonically increasing with black hole spin $a$ and
the scattered (high energy) photons are more abundant. 

All our models are anisotropic sources of radiation and equatorial
observers are preferred in all cases. The effect is not as strong as
described in Paper I, but not negligible. The largest anisotropy is present
in low accretion rate (${\dot m}=10^{-3}$) ADAFs around Kerr ($a=0.9$)
black holes. It must be caused by the relativistic effects in light
scattering and propagation (mostly the Doppler effect).

\Acknow{This work was supported in part by the Polish State Committee
for Scientific Research grants 2-P03D-012-16 and 2-P03D-013-16}


\begin{references}

  \refitem{Abramowicz, M.A., Chen, X., Granath, M., and Lasota, J.-P.}
          {1996}{Astrophys. J.}{471}{762}

  \refitem{Abramowicz, M.A., Lanza A., and Percival M.J.}
          {1997}{Astrophys. J.}{479}{179}

  \refitem{Abramowicz, M.A., Lasota, J.-P., and Igumenshchev, I.V.}
          {2000}{astro-ph/0001479}{~}{~}

  \refitem{Bardeen, J.M.}{1973}{In: Black Holes}{~}{ed. C. DeWitt
        and B.S. DeWitt (Gordon \& Breach, New York), p. 215}

  \refitem{Blandford, R.D., and Begelman, M.C.}{1999}{MNRAS}{303}{L1} 

  \refitem{Gammie, C.F., and Popham, R.G.}{1998}{Astrophys. J.}{498}{313}

  \refitem{Jaroszy\'nski, M., and Kurpiewski, A.}
          {1997}{Astron. Astrophys.}{326}{419 (Paper I)}

  \refitem{Kato, S., Fukue, J., and Mineshige, S.}{1998}
          {Black Hole Accretion Disks}{~}{(Kyoto University Press, Kyoto)}

  \refitem{Kurpiewski, A., and Jaroszyñski, M.}{1999}{Astron. Astrophys.}
          {346}{713 (Paper II)}

  \refitem{Lasota, J.-P.}{1994}{In: Theory of the Accretion
  	Disks 2}{~}{ed. W.J. Duschl, J. Frank, F. Meyer, E.
        Meyer-Hoffmeister 
	and W.M. Tscharnuter (Kluwer Academic Publishers, 
	Dordrecht), p. 341}

  \refitem{Lasota, J.-P.}{1999}{Phys. Reports}{311}{247}

  \refitem{Nakamura, K.E., Kusunose, M., Matsumoto, R., and Kato, S.}
          {1997}{P.A.S.J.}{49}{503}

  \refitem{Narayan, R., and Yi, I.}{1994}{Astrophys. J.}{428}{L13}

  \refitem{Narayan, R., and Yi, I.}{1995}{Astrophys. J.}{444}{231}

  \refitem{Narayan, R., Mahadevan, R., Grindlay, J.E., Popham, R.G., and
           Gammie, Ch.}{1998}{Astrophys. J.}{492}{554 (NMGPG)}
          
  \refitem{Narayan, R., Mahadevan, R., and Quataert, E.}{1998}
          {In: The Theory of Black Hole Accretion Disks}{~}{ed. 
           M.A. Abramowicz, G. Bj\"ornsson and J. Pringle
           (Cambridge University Press, Cambridge)}

  \refitem{Paczy\'nski, B.}{1998}{Acta Astron.}{48}{667}

  \refitem{Park, M.-G., and Ostriker, J.P.}{1999}{Astrophys. J.}{527}{247}

  \refitem{Park, M.-G., and Ostriker, J.P.}{2000}{astro-ph/0001446}{~}{~}

  \refitem{Peitz, J., and Appl, S.}{1997}{MNRAS}{286}{681}

  \refitem{Popham, R.G., and Gammie, C.F.}{1998}{Astrophys. J.}{504}{419}

  \refitem{Svensson, R.}{1998}{In: Theory of Black Hole Accretion Disks}
      {~}{ed. M.A. Abramowicz, G. Bj\"ornson and J.E. Pringle  
      (Cambridge University Press)}

  \refitem{Yi, I.}{1999}{In: Astrophysical Discs, ASP Conf. Series 160, 
          astro-ph/9905215}{~}{~}

\end{references}
\end{document}